\documentclass[aps,preprint]{revtex4-1}
\usepackage{graphicx}
\usepackage{amssymb,amsmath,amsbsy,amsfonts,bm}
\usepackage{epsfig}
\usepackage[percent]{overpic}
\usepackage{color}

\begin{document}
\title{Anomalous transport of magnetic colloids in a liquid crystal-magnetic colloid mixture}

\author{Gaurav P. Shrivastav}
\email{shrivastav@tu-berlin.de}
\affiliation{Institut f\"ur Theoretische Physik, Technische Universit\"at 
Berlin, Hardenberg Str. 36, 10623 Berlin, Germany.}

\author{Sabine H. L. Klapp}
\email{klapp@physik.tu-berlin.de}
\affiliation{Institut f\"ur Theoretische Physik, Technische Universit\"at 
Berlin, Hardenberg Str. 36, 10623 Berlin, Germany.}

\begin{abstract}
We report an extensive molecular dynamics study on the translational dynamics of a hybrid system composed of dipolar soft spheres (DSS), representing ferromagnetic particles, suspended in a liquid crystal (LC) matrix. We observe that the LC matrix strongly modifies the dynamics of the DSS. In the isotropic regime, the DSS show a crossover from subdiffusive to normal diffusive behavior at long times, with an increase of the subdiffusive regime as the dipolar coupling strength is increased. In the nematic regime, the LC matrix, due to collective reorientation of LC particles, imposes a cylindrical confinement on the DSS chains. This leads to a diffusive dynamics of DSS along the nematic director and a subdiffusive dynamics (with an exponent $\sim 0.5$) in the perpendicular direction. The confinement provided by the LC matrix is also reflected by oscillatory behavior of the components of the velocity autocorrelation function of the DSS in the nematic phase.
\end{abstract}
\maketitle
\section{Introduction}
Inclusions of nano-to-micro sized particles inside a liquid crystal (LC) matrix have offered a new paradigm for hybrid material design and applications of LCs beyond display materials \cite{ls16,ls12,bk11}. This is one of most propitious advances in the material science as the self-organizing tendency of LCs provides numerous possibilities to synthesize fascinating bulk structured materials \cite{bcl13}. In this context, suspensions of magnetic particles inside a LC matrix have attracted particular attention in the past few decades. These systems, first introduced theoretically in a celebrated work by Brochard and de Gennes \cite{gb70}, show a rich variety of self-assembled structures and have a wide range of biomedical and technical applications \cite{lms09,sn05}. The first experimental realization of such suspensions was achieved by doping magnetic particles in a thermotropic LC host \cite{rcb70}. A further early experimental study was by L\'{e}bert and Martinet who doped lyotropic LC with a water-based ferrofluid and observed that the magnetic field required to align LC is reduced by a factor of thousand \cite{lm79}. 

Moreover, these materials provide striking magneto-optical properties due to the combination of the anisotropic optical properties of LCs and the large magnetic susceptibility of the magnetic particles. Many experiments have investigated the effects of external magnetic field on the host LC matrix \cite{ns86,ca83,ktkz08,bbc98,pbb11,bnr11,kbd13,mldc13,ao15}. In particular, in a recent combined experimental and numerical study, it has been shown that the presence of magnetic particles, combined with an external magnetic field, can induce a nematic phase in an isotropic LC phase \cite{mespkk16}.

From a theoretical point of view, a standard model for spherical ferromagnetic particles are so-called dipolar soft spheres (DSS), that is, spheres interacting via a steep isotropic repulsion and anisotropic forces stemming from embedded point dipoles. The equilibrium structure of DSS in a LC matrix composed of Gay-Berne ellipsoids are well studied theoretically \cite{br95,rpm03,jpb03} and via computer simulations for different sizes of the two species \cite{plk15,pk15,pk16}. In particular, Monte Carlo (MC) simulations reveal that in the absence of an external magnetic field the DSS form chains along the director of the nematic LC matrix. Further, as the diameter of DSS is increased, the uniaxial spontaneous ordering changes to a biaxial lamellar phase \cite{pk15}. 

In order to control and improve the flow properties of such hybrid systems \cite{od03}, it is important to develop a better understanding of the equilibrium dynamics of different species, in particular, the translational mobility. In a recent molecular dynamics (MD) simulations \cite{pk16}, it has been observed that the DSS show normal diffusion if their sizes are much smaller than the suspending LC fluid. Nonetheless, the dynamics may become more complex when the two components have comparable sizes. In fact, dilute dipolar fluids in the absence of a LC matrix display a subdiffusive behavior at intermediate times due to the formation of chains with head-to-tail ordering of the dipole moments \cite{jss11}. This sudiffusive regime increases as the strength of dipolar coupling is increased. Also, at a dipolar coupling strength where dipolar particles start to form chains, the translational diffusion coefficient shows a sudden decrease \cite{si13}. Thus, the chain-forming tendency of dipolar particles plays an important role on their dynamics. We would therefore expect that the situation gets even more complex in a hybrid system of DSS in a LC host. Here, the LC matrix provides a dense anisotropic environment which, as we will show, renders the dynamics of magnetic particles anomalous. This finding is indeed consistent with the behavior of a wide range of complex systems where a dense environment or a complex geometry leads to a subdiffusive motion of particles \cite{hf13,bgm12,ms15,hss17}. 

In this work, we present a MD simulation study on the translational dynamics of a mixture of LC and DSS, where the sizes of both the species are comparable. We observe that the dynamics of DSS is indeed highly influenced by the LC matrix. At low densities where the LC matrix is in an isotropoic phase, the mean square displacement (MSD) of DSS shows a subdiffusive regime at intermediate times. The length of the subdiffusive regime grows as the strength of dipolar coupling is increased. Eventually it spans the entire simulation time window at high values of the dipolar coupling. The effect of the LC matrix on the dynamics of the DSS is particularly visible at high densities, where the LC matrix undergoes an I-N transition and enforces, in turn, an aligned state of the DSS \cite{pk15}. The DSS then show diffusive behavior parallel to the nematic director but remain subdiffusive in the perpendicular direction.

The rest of the paper is organized as follows. In Sec.~\ref{sd} we give the details of model and simulation method. We present our results in sec.~\ref{res}. The equilibrium phase diagram of the mixture is discussed in sec.~\ref{pd}. Using mean square displacements (MSD) (Sec.\ref{msd}) and velocity autocorrelation functions (VACF) (Sec.~\ref{vacf}), we demonstrate that the DSS show anomalous translational diffusion at low temperatures while the LC matrix shows a normal diffusive behavior at all densities and temperatures. Finally, in Sec.~\ref{con}, we conclude the paper with a summary. 

\section{Simulation details}
\label{sd}
We consider a binary mixture of LC and DSS with a composition ratio 80:20 and perform MD simulations in the NVT ensemble using the LAMMPS package \cite{pl95,bpp09}. The LCs are modeled by ellipsoids which are characterized via a (diagonal) shape matrix ${\bm S} = {\rm diag}\left(\sigma_{a}, \sigma_{b}, \sigma_{c}\right)$ and a (diagonal) energy matrix ${\bm E} = {\rm diag}\left(\epsilon_{a}, \epsilon_{b}, \epsilon_{c}\right)$, where $\sigma_{a, b, c}$ are the lengths and $\epsilon_{a, b, c}$ are the relative well depths of interaction along the three semiaxes of an ellipsoid. We note that the model investigated here slightly differs from that considered in Peroukidis et. {\it al}. \cite{pk16} as the mixing parameters are different.

The LCs interact via a generalized Gay-Berne (GB) potential which is defined as \cite{bpp09,bcm07,ee03} (following the notations of Brown et. {\it al.} \cite{bpp09}) 
\begin{eqnarray}
{\rm U}\left({\bm A}_{i},{\bm A}_{j}, {\bm r}_{ij}\right) = {\rm U}_{r}\left({\bm A}_{i},{\bm A}_{j}, \bm{r}_{ij}\right)\eta_{ij}\left({\bm A}_{i}, {\bm A}_{j}\right)\chi_{ij}\left({\bm A}_{i}, {\bm A}_{j}, \bm{\hat{r}_{ij}}\right).
\label{gb}
\end{eqnarray}
Here ${\bm A}_{i}$ is the rotation matrix for a particle $i$, used for the transformation from the lab frame to the body frame of reference. Further, $\bm{r}_{ij}$ is the center-to-center distance vector between particles $i$ and $j$, and $\bm{\hat{r}_{ij}}$ is the unit vector along $\bm{r}_{ij}$. 

The function ${\rm U}_{r}\left({\bm A}_{i},{\bm A}_{j},\bm{r}_{ij}\right)$, which controls the distance dependence of the GB potential, is defined as
\begin{eqnarray}
{\rm U}_{r}\left({\bm A}_{i}, {\bm A}_{j}, {\bm r}_{ij}\right) = 4\epsilon_{0}\left[\left(\frac{\sigma_{0}}{h_{ij}+\gamma \sigma_{0}}\right)^{12} - \left(\frac{\sigma_{0}}{h_{ij}+\gamma \sigma_{0}}\right)^{6} \right],
\label{gbr}
\end{eqnarray}
where $\epsilon_{0}$ and $\sigma_{0}$ set the units of energy and length, $\gamma$ is the shift parameter and $h_{ij}= r_{ij} - \left[\frac{1}{2}\hat{r}^{T}_{ij}{\bm G}_{ij}\hat{r}_{ij}\right]^{-1/2}$ is the distance of closest approach between particles $i$, $j$ with ${\bm G}_{ij} = {\bm A}^{T}_{i}{\bm S}^{2}_{i}{\bm A}_{i} + {\bm A}^{T}_{j}{\bm S}^{2}_{j}{\bm A}_{j}$.
The second and third term in Eq.~(\ref{gb}) are given by
\begin{eqnarray}
\label{eta}
\eta_{ij}\left({\bf A}_{i}, {\bm A}_{j}\right) &=& \left[\frac{2s_{i}s_{j}}{det\left[{\bm G}_{ij}\left({\bm A}_{i}, {\bm A}_{j}\right)\right]}\right]^{\nu/2},\\
\chi_{ij}\left({\bm A}_{i}, {\bm A}_{j}, \hat{\bm r}_{ij}\right) &=& \left[2\hat{\bm r}^{T}_{ij}{\bm B}^{-1}_{ij}\left({\bm A}_{i}, {\bm A}_{j}\right)\hat{\bm r}_{ij}\right]^{\mu^{\prime}}.  
\label{gen}
\end{eqnarray}
In Eq.~(\ref{eta}), $s_{i,j} = \left[\sigma_{a_{i,j}}\sigma_{b_{i,j}} + \sigma_{c_{i,j}}\sigma_{c_{i,j}}\right]\left[\sigma_{a_{i,j}}\sigma_{b_{i,j}}\right]^{1/2}$ and ${\bm B}_{ij} = {\bm A}^{T}_{i}{\bm E}^{2}_{i}{\bm A}_{i} + {\bm A}^{T}_{j}{\bm E}^{2}_{j}{\bm A}_{j}$ with ${\bm S}_{i,j}$ and ${\bm E}_{i,j}$ representing the shape and energy matrices for particles $i$ and $j$.

We consider uniaxial LCs with aspect ratio 3, i.e., $\sigma^{e}_{a} = \sigma^{e}_{b} = \sigma_{0}$ and $\sigma^{e}_{c} = 3 \sigma_{0}$. The relative energy well depths for side-to-side interaction is $\epsilon^{e}_{a} = \epsilon^{e}_{b} = \epsilon_{0}$ and for end-to-end interaction is $\epsilon^{e}_{c} = 0.2\epsilon_{0}$. For ellipsoids, these values of the energy parameters yield the energy of side-to-side configuration five times stronger than end-to-end configuration. The cut-off radius is set to $r^{\rm GB}_{c} = 4.0\sigma_{0}$ and the empirical parameters of GB potential are set to $\mu^{\prime} = 1.0, \nu = 2.0$ and $\gamma = 1.0$. The parameter $\mu^{\prime}$ is same as $\mu$ used for GB in Brown et. {\it al.} \cite{bpp09}. Here, we have changed the notation, as in our case, $\mu$ is reserved for the dipole moment of the DSS.

The interaction among the LC and DSS is also modeled by a GB potential with shape and energy parameters for DSS taken as $\sigma^{s}_{a} = \sigma^{s}_{b} = \sigma^{s}_{c} = \sigma_{0}$ and $\epsilon^{s}_{a} = \epsilon^{s}_{b} = \epsilon^{s}_{c} = \epsilon_{0}$. The cutoff radius is taken same as $r^{\rm GB}_{c}$. 

The DSS interact via a combination of a soft sphere potential and dipolar interactions \cite{wp92,mdk06}. The full potential for two DSS particles $i$ and $j$ with dipole moments $\bm{\mu}_{i}$ and $\bm{\mu}_{j}$ is defined as \cite{mdk06}
\begin{eqnarray}
{\rm U}(ij) = {\rm U}_{\rm SR}\left(r_{ij}\right) + \frac{\bm{\mu}_{i}\cdot \bm{\mu}_{j}}{r^{3}_{ij}} - 3\frac{\left({\bm{\mu}}_{i}\cdot {\bm{r}}_{ij}\right)\left({\bm{\mu}}_{j}\cdot {\bm{r}}_{ij}\right)}{r^{5}_{ij}},
\label{sst}
\end{eqnarray}
where ${\rm U}_{\rm SR}\left(r_{ij}\right)$ is the shifted-force soft sphere interaction given as
\begin{eqnarray}
{\rm U}_{\rm SR}\left(r_{ij}\right) = {\rm U}_{\rm SS}\left(r_{ij}\right) - {\rm U}_{\rm SS}(r^{\rm SS}_{c}) - \left(r^{\rm SS}_{c} - r_{ij}\right)\frac{d{\rm U}_{\rm SS}}{dr}\Bigg\vert_{r = r^{\rm SS}_{c}}.
\label{ssr}
\end{eqnarray}
In Eq.~(\ref{ssr}), the cutoff radius for soft sphere potential is set to $r^{\rm SS}_{c} = 2.5\sigma_{0}$, and 
\begin{eqnarray}
{\rm U_{\rm SS}}\left(r_{ij}\right) = 4\epsilon_{0}\left(\sigma_{0}/r_{ij}\right)^{12}.
\label{ss}
\end{eqnarray}
The long range dipolar interactions are treated with the three dimensional Ewald sum \cite{sk07,whm02}. 

The parameters that characterize the structure and phase behavior of the mixture are the reduced temperature ${\rm T}^{*} = k_{\textrm B}T/\epsilon_{0}$, the reduced number density $\rho^{*} = N \sigma_{0}^{3}/V$ (where $N$ and $V$ are the total number of particles and total volume respectively), and the reduced dipole moment $\mu^{*} = \mu^{2}/\epsilon_{0}\sigma^{3}$. The Newton's equations of motion for force and torque are integrated in the NVT ensemble using velocity Verlet algorithm. A reduced MD time step $\Delta t^{*} = \Delta t/\sqrt{m\sigma_{0}^{2}/\epsilon_{0}} = 0.002$ is chosen and simulations are performed at fixed $\mu^{*} = 3.0$, various $\rho^{*}$ and $\rm T^{*}$. The value of $\rm T^{*}$ ranges from 2.0 to 0.6, i.e., the range of dipolar coupling parameter $\lambda = \mu^{2}/k_{\textrm B}T\sigma_{0}^{3}$ is 4.5 to 15.0. We have set $\mu^{*} = 3.0$, therefore, a change in ${\rm T}^{*}$ is equivalent to a change in $\lambda$. We will use $\lambda$ for DSS and the corresponding ${\rm T}^{*}$ for LCs while discussing our results. 

Our simulated system consists of 3200 LC ($N_{e}$) and 800 DSS particles ($N_{s}$). We start with a mixture equilibrated at low density and high temperature and then quench it to desired $\rm T^{*}$. Subsequently, we slowly compress the mixture keeping $\rm T^{*}$ constant by applying Langevin thermostat \cite{at06}. At each compression step, the mixture is equilibrated for $4\times 10^{6}$ time steps. For MSD calculations at different $\rho^{*}$, we performed additional production runs in the NVT ensemble for $10^{7}$ time steps after equilibration. We took 100 time origins for averaging. 

The self-diffusion constant $D$ is obtained from the long-time behavior of MSD. For VACF calculations, short production runs up to 25000 time steps are performed using temperature rescaling at a frequency of 1000 time steps. It should be noted that performing long production runs in the NVE ensemble is quite difficult for the mixture. Also, the Langevin thermostat affects the equilibrium fluctuations by applying random forces on the particles. Therefore, we use temperature rescaling during the VACF production runs in order to maintain the temperature. The timescale for temperature rescaling is chosen such that the VACF decays almost to zero in that time.
\section{Results}
\label{res}
In this section we first present our numerical results for the equilibrium phase diagram of the LC-DSS mixture at the selected dipole moment $\mu = 3.0$. Second, we discuss the translational dynamics of the two components. In particular, we investigate the respective MSDs and normalized velocity autocorrelation functions (VACF). 
\subsection{Equilibrium phase diagram}
\label{pd}
First, we identify the I-N transition by calculating the nematic order parameter $S$ for the LC-DSS mixture. This order parameter is defined as the largest eigenvalue of the ordering tensor ${\bm Q}$. The components of the $\bm Q$-tensor are given by $Q_{\alpha\beta} = \left(1/N \right)\sum_{i=1}^{N}(1/2)\left(3\hat{u}_{\alpha}^{i}\hat{u}_{\beta}^{i}-\delta_{\alpha\beta}\right)$. Here, $\alpha, \beta = x, y, z$ and $\hat{u}^{i}$ is, for LC, the orientation vector and for DSS, the unit dipole vector $\hat{\mu}^{i}$. The nematic order parameters $S_{e}$ and $S_{s}$ for individual components, LC and DSS, are calculated separately using respective $\bm Q$ tensor. The eigenvectors corresponding to $S_{e}$ and $S_{s}$ define the directors $\hat{n}_{e}$ and $\hat{n}_{s}$ for LC and DSS respectively. 

\begin{figure}
 \centering
 \includegraphics[scale=0.6]{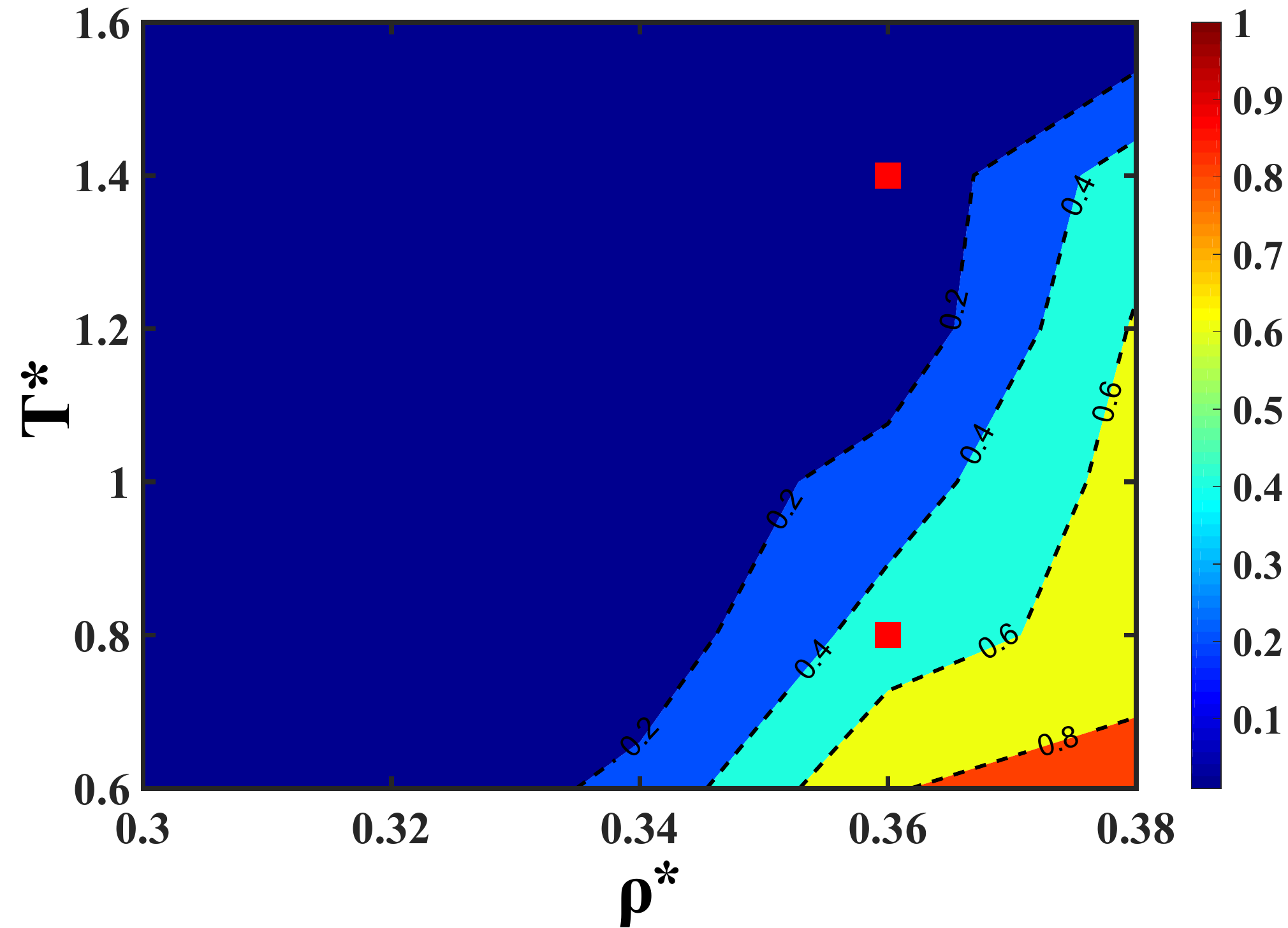}
 \caption{The $\rho^{*}-\rm T^{*}$ phase diagram of the LC matrix. The color axis shows the nematic order parameter $S_{e}$, and dashed lines mark the contours of different $S_{e}$ values. The dashed line corresponding to $S_{e} = 0.4$ is considered as the tentative I-N phase transition line. Red squares represent state points $(0.36, 1.4)$ and $(0.36, 0.8)$ in the isotropic and nematic phase, respectively. Snapshots corresponding to these two state points are shown in Fig.~\ref{fig1c}.}
 \label{fig1a}
\end{figure}

\begin{figure}
 \centering
 \includegraphics[scale=0.6]{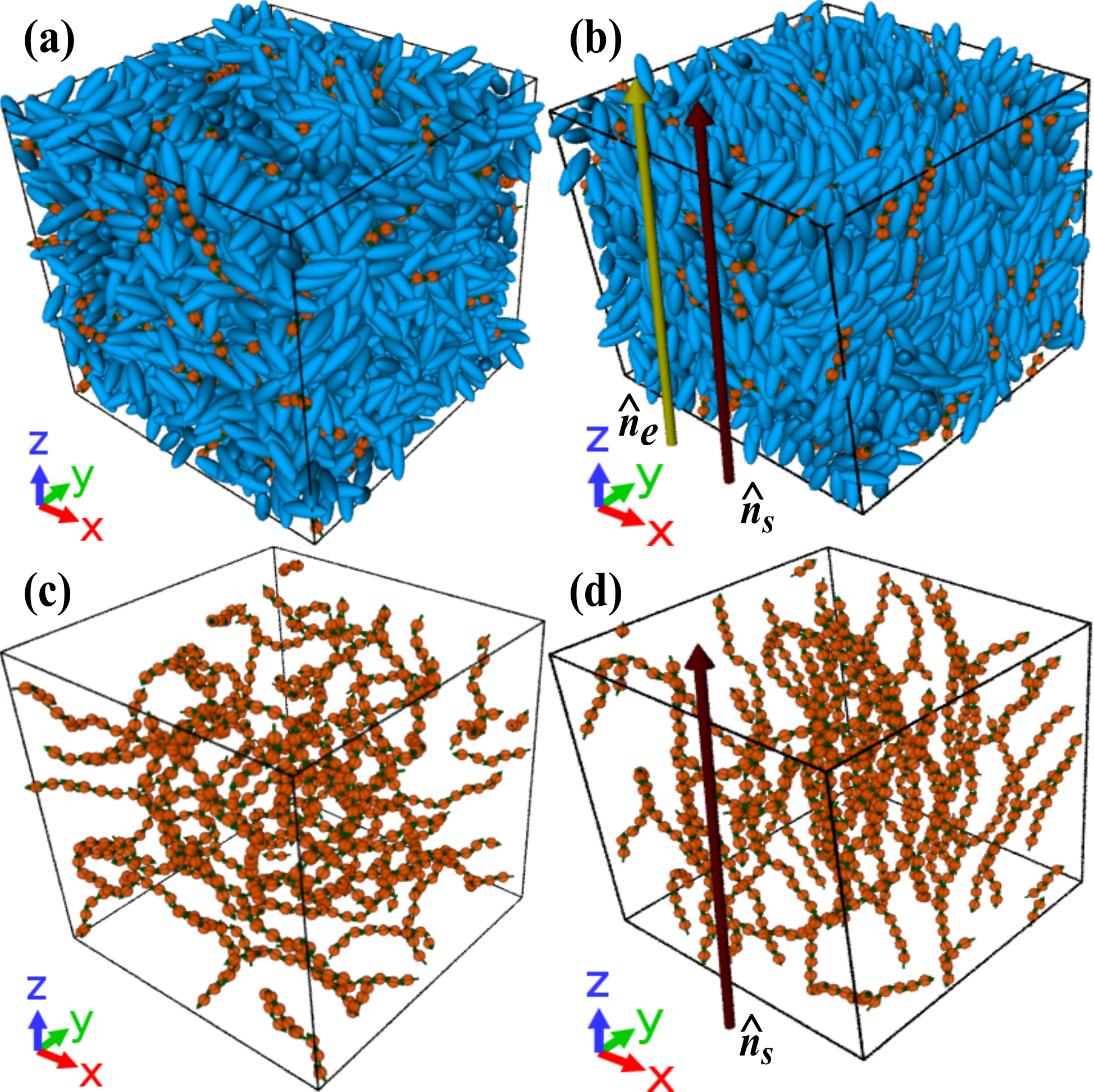}
 \caption{Snapshots of the LC-DSS mixture at the state points marked by the red squares in the Fig.~\ref{fig1a}. (a) snapshot of the LC-DSS mixture at $\rho^{*} = 0.36$ and $\rm T^{*} = 1.4$, (b) snapshot of the LC-DSS mixture at $\rho^{*} = 0.36$ and $\rm T^{*} = 0.8$. (c) and (d) show the DSS chains in the absence of the LC matrix for the state points considered in (a) and (b), respectively. All the snapshots are prepared using software OVITO \cite{ov10}.}
 \label{fig1c}
\end{figure}

In Fig.~\ref{fig1a}, we plot $S_{e}$ for the LC matrix (shown by the color axis) as a function of $\rho^{*}$ and $\rm T^{*}$. At low $\rho^{*}$ and high $\rm T^{*}$, the LC matrix remains in an isotropic phase while it undergoes an I-N transition at high $\rho^{*}$ and low $\rm T^{*}$ values. The DSS chains follow the LC matrix and undergo an I-N transition at the same values of $\rho^{*}$ and $\rm T^{*}$ (not shown here but visible in the snapshots in Fig.~\ref{fig1c}). The dashed lines in the plot represent the contours of different $S_{e}$ values. We consider the line corresponding to $S_{e} = 0.4$ as the tentative I-N transition line. For illustration of the actual structure, snapshots of the mixture corresponding to two state points (marked in red squares) are plotted in Fig.~\ref{fig1c}. At $\rho^{*} = 0.36$ and $\rm T^{*} = 1.4$, the mixture is in an isotropic phase. As shown in Fig.~\ref{fig1c}(a), the LCs and the DSS chains (shown in Fig.~\ref{fig1c}(c)) are randomly oriented. Further, Fig.~\ref{fig1c}(b) and (d) show that at $\rho^{*} = 0.36$ and $\rm T^{*} = 0.8$, the LC matrix and DSS chains are both in a nematic phase. Also, the two nematic directors $\hat{n}_{e}$ and $\hat{n}_{s}$ are aligned almost parallel to each other which is consistent with the previous MC simulations \cite{plk15,pk15}.

\subsection{Mean-square displacements}
\label{msd}
We, next, analyze the translational dynamics of the two components of the mixture using MSD which is defined as 
\begin{eqnarray}
\label{msdeq}
\langle\Delta r^{2}\rangle = \frac{1}{N_{e,s}}\sum_{i=1}^{N_{e,s}}\langle\left|\bm{r}_{i}(t+t_{0}) - \bm{r}_{i}(t_{0})\right|^2\rangle,
\end{eqnarray}
where $\bm{r}_{i}(t)$ is the position of the particle at time $t$, $t_{0}$ is the time origin for MSD calculations, and angular bracket corresponds to the averaging over the total number of particles of a species, time origins and number of samples. To extract the longtime behavior of MSD, we calculate the instantaneous slopes, $\phi(t)$, of the MSD curves which is defined as 
\begin{eqnarray}
\label{msdslope}
\phi(t) = d\{{\rm ln}(\left\langle\Delta r^{2}\rangle\right)\}/d\{{\rm ln}(t)\}.
\end{eqnarray}
The slope of MSD curves at long times is given by 
\begin{eqnarray}
\label{slopemsd}
\alpha = \lim_{t\to \infty} \phi(t).
\end{eqnarray}

In the nematic phase, displacements of particles are resolved in the components parallel, ${\Delta \bm{r}_{||}}$, and perpendicular, ${\Delta \bm{r}_{\perp}}$, to the LC/DSS nematic director $\hat{n}_{e,s}$ at $t_{0}$ and corresponding MSDs \big($\langle \Delta r_{||}^{2}\rangle$ and $\langle \Delta r_{\perp}^{2}\rangle$\big) are calculated \cite{hl99,mrg92}. For LCs we extract diffusion coefficients using Stokes-Einstein relation, ${\rm D} = \left(1/6t\right)lim_{t \to \infty} \langle\Delta r^{2}\rangle$. The diffusion constants in the directions parallel and perpendicular to the $\hat{n}_{e}$ are defined as 
\begin{eqnarray}
\label{diffcoef}
{\rm D_{||, \perp}} = \left(1/2t\right)lim_{t \to \infty} \langle\Delta r^{2}_{||, \perp}\rangle.
\end{eqnarray}
\subsubsection{Diffusional behavior of the dipolar particle}
\begin{figure}
 \centering
 \includegraphics[scale=0.6]{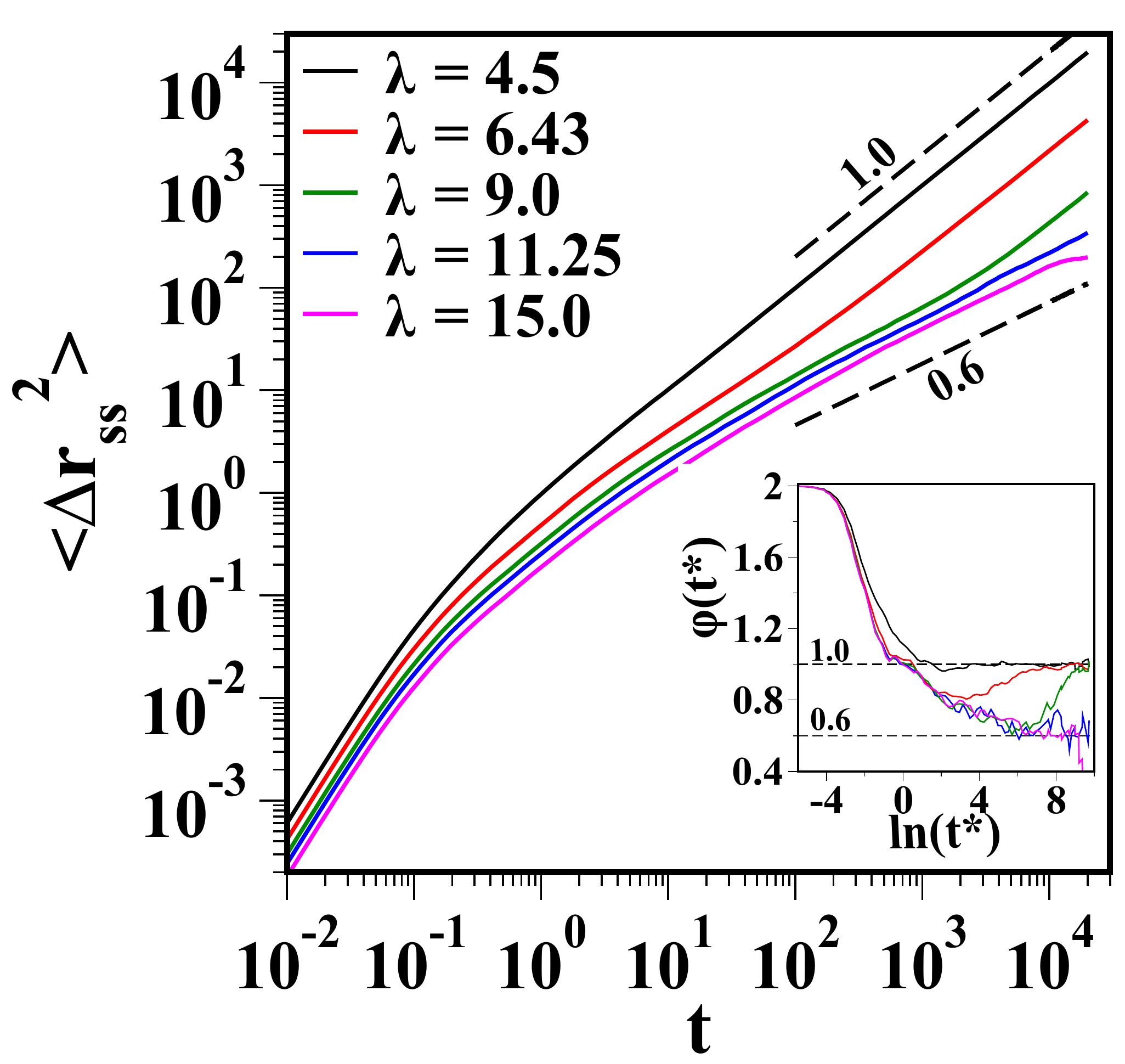}
 \caption{(a). MSD of DSS in the isotropic regime for $\rho^{*} = 0.3$ at $\lambda = 4.5, 6.43, 11.25, 15.0$. At low $\lambda$, the MSD has a slope of 1.0 at large $t$ (shown by black dashed line). At large $\lambda$, subdiffusive behavior is observed with an exponent 0.6. The inset shows the instantaneous slope $\phi$ of MSDs, which saturates to 1.0 for low $\lambda$ while to 0.6 for large $\lambda$ at long times.}
 \label{fig1}
\end{figure}

The MSDs of the DSS in the isotropic phase are plotted in Fig.~\ref{fig1} for $\rho^{*} = 0.3$ and various values of $\lambda$. For all $\lambda < 11.25$, the MSDs show three different regimes, ballistic at short times, subdiffusive at intermediate times and diffusive at long times. At low $\lambda$, the subdiffusive regime is rather short, however, as $\lambda$ is increased the subdiffusive regime grows, and for $\lambda > 9.0$ at $\rho^{*} = 0.3$, it spans the entire simulation time window. The crossover from diffusion to subdiffusion at long times with increasing $\lambda$ is more evident in the inset of Fig.~\ref{fig1}. Here $\phi(t)$ is plotted for different MSD curves (shown in Fig.~\ref{fig1}) as a function of $t^{*}$ at $\rho^{*} = 0.3$. Clearly, at low $\lambda$, $\phi(t)$ saturates to 1 while it saturates to 0.6 at higher $\lambda$ at long times. A region with slope less than 1 exists at all $\lambda > 4.5$ which grows with increasing $\lambda$ and eventually extends over the entire simulation time window. 

\begin{figure}
 \centering
 \includegraphics[scale=0.6]{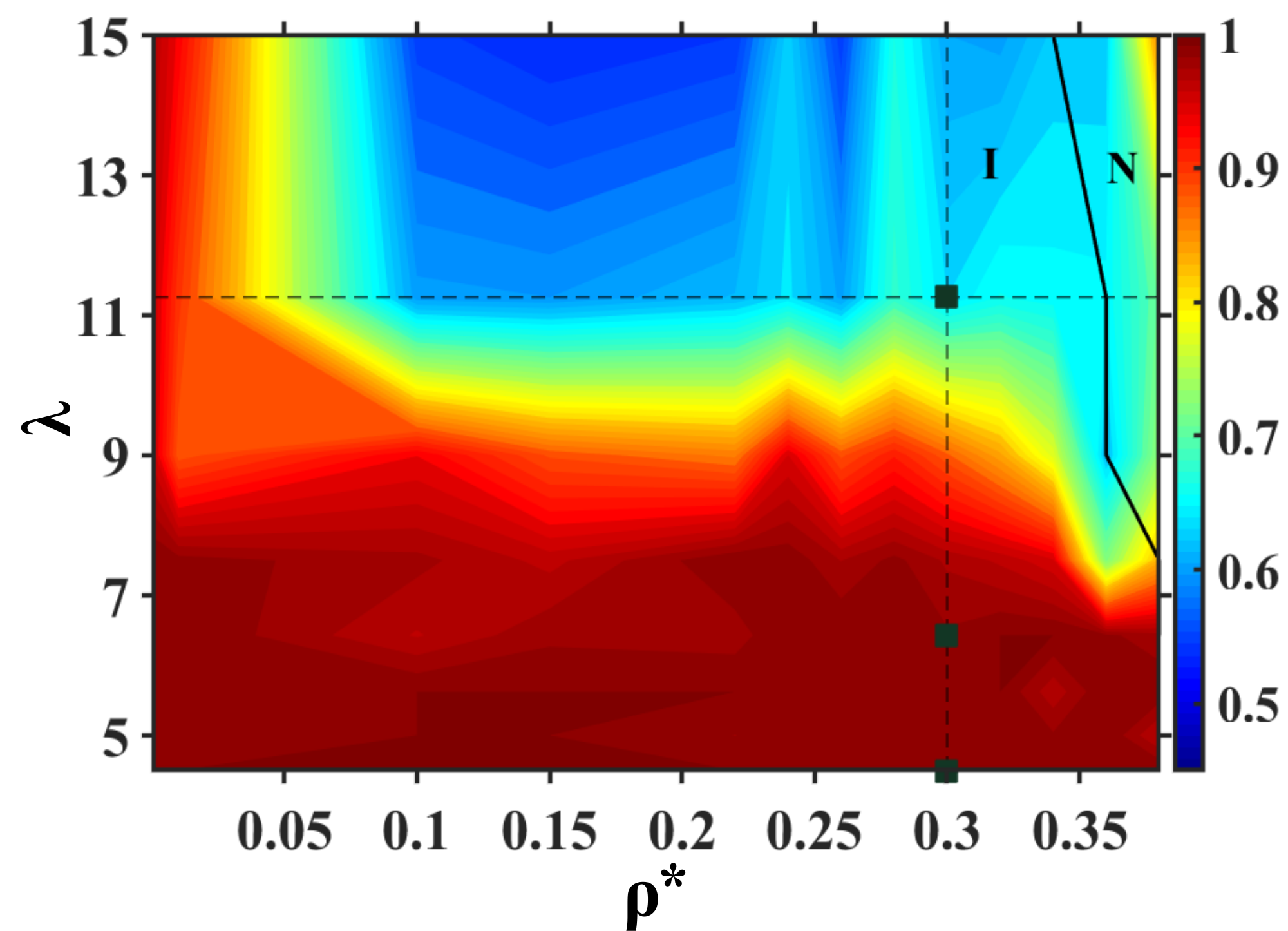}
 \caption{Contour map showing $\alpha$ (see Eq.~\ref{slopemsd}), extracted from the long-time MSD of DSS, as a function of $\lambda$ (corresponding $\rm T^{*}$ can be obtained using $\rm T^{*} = 9.0/\lambda$) and $\rho^{*}$. One observers a subdiffusive regime at large $\lambda$ in the isotropic phase. The black solid line represents the I-N transition. The vertical and horizontal dashed lines correspond to fixed $\rho^{*} (= 0.3)$ and fixed $\lambda (= 11.25)$, respectively. The green squares, marked on the $\rho = 0.3$ (vertical) line, show three state points $\lambda = 4.5, 6.43$ and $\lambda = 11.25$ at $\rho^{*} = 0.3$.}
 \label{fig2}
\end{figure}

In order to understand the longtime behavior of the MSD of the DSS at different $\rho^{*}$ and $\lambda$, we have obtained a state diagram of $\alpha$ as a function of $\rho^{*}$ and $\lambda$, see Fig.~\ref{fig2}. Here, the color axis represents $\alpha$ and a black solid line shows the I-N transition, which appears at high $\rho^{*}$ and $\lambda$. For $\lambda < 7.0$, $\alpha$ remains close to 1.0 for all considered $\rho^{*}$, reflecting normal diffusive behavior of MSDs. As $\lambda$ is increased, the longtime behavior becomes subdiffusive at high densities and for $\lambda > 11.0$, the dynamics of DSS appears to be subdiffusive even at lower densities. 

For a better understanding of the crossover of the longtime behavior of MSDs, we explore $\alpha$ along the two black dashed lines shown in the Fig.~\ref{fig2}. The vertical dashed line represents fixed $\rho^{*}$ (=0.3) and different $\lambda$, while the horizontal dashed line corresponds to fixed $\lambda$ (=11.25) and different $\rho^{*}$s. 

\begin{figure}
 \centering
 \includegraphics[scale=0.8]{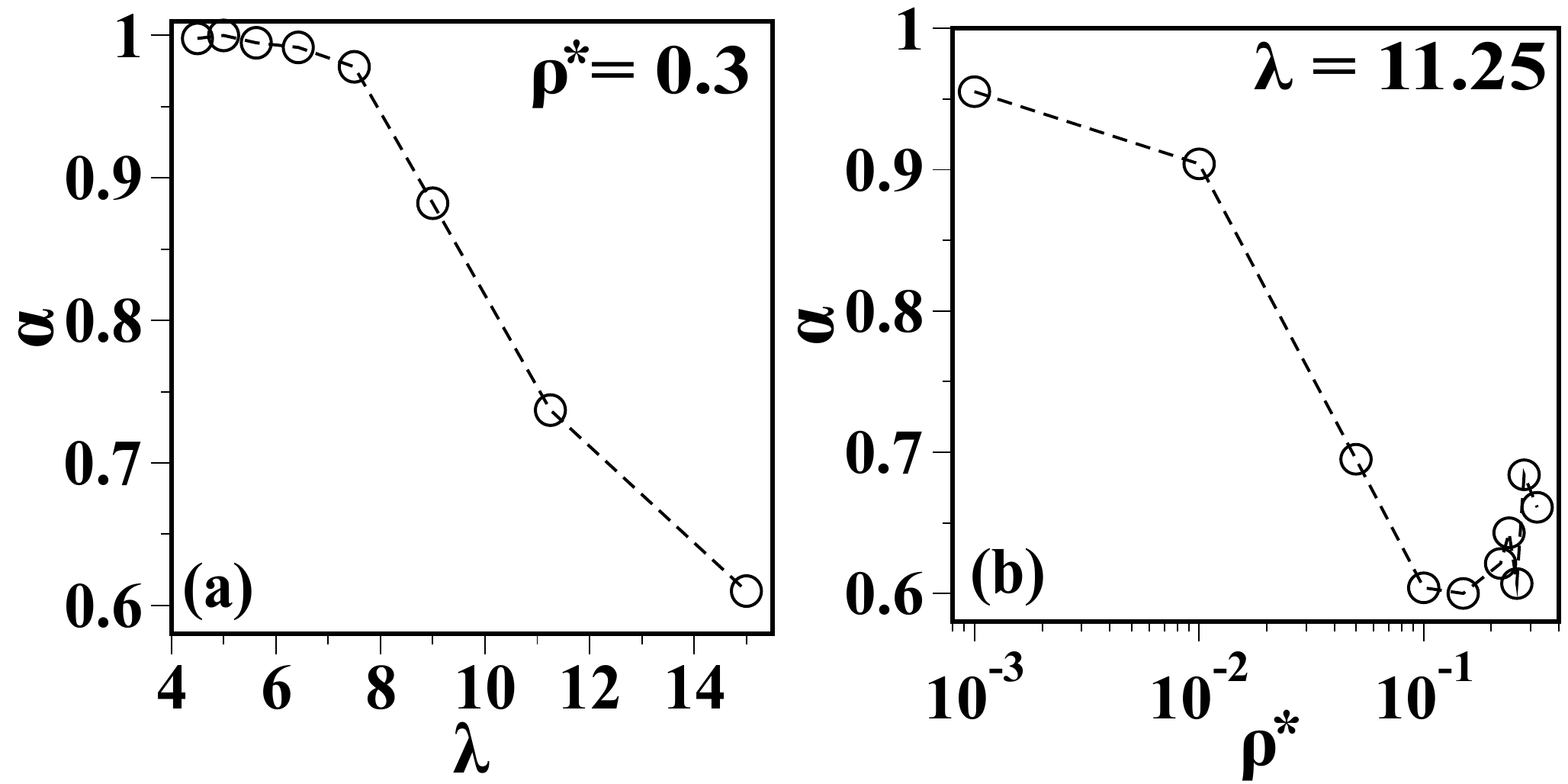}
 \caption{Longtime slopes, $\alpha$, of MSDs of DSS, (a) Data for $\rho^{*} = 0.3$ as a function of $\lambda$ and (b) data for $\lambda = 11.25$ as a function of $\rho^{*}$.}
 \label{fig3}
\end{figure}

The variation of $\alpha$ as a function of $\lambda$ at a fixed $\rho^{*} =0.3$ (along the vertical dashed line in Fig.~\ref{fig2}) is shown in Fig.~\ref{fig3}(a). Clearly, $\alpha$ shows a crossover from $1.0$ to $0.6$ as $\lambda$ is increased. A similar crossover (Fig.~\ref{fig3}(b)) is observed when $\rho^{*}$ is varied at  fixed $\lambda = 11.25$. However, Fig.~\ref{fig2} suggests that the $\lambda$ has to be large ($> 7.0$) in order to see a crossover as a function of $\rho^{*}$. 

\begin{figure*}
 \centering
 \includegraphics[scale=0.8]{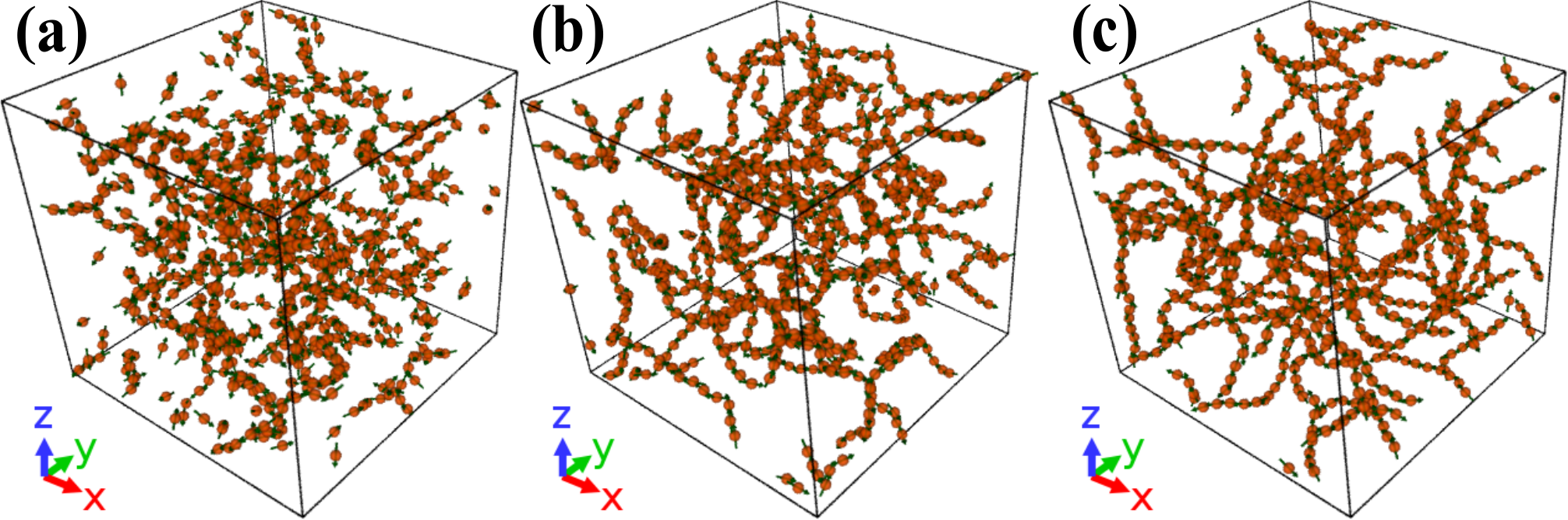}
 \caption{(a). Snapshots showing DSS chains in the absence of the LC matrix at (a) $\lambda = 4.5$, (b) $\lambda = 6.43$, (c) and $\lambda = 11.25$ for $\rho^{*} = 0.3$, (see the green square points marked in the Fig.~\ref{fig2}). At these values of $\lambda$ and $\rho^{*}$, the mixture always remains in the isotropic phase, however, the length of the DSS chains increases as $\lambda$ is increased.}
 \label{fig4}
\end{figure*}

In the case of pure DSS, at $\rho^{*} = 0.05$, $\lambda = 7.0$ and in the absence of an external magnetic field, a sublinear regime in the MSDs is observed at intermediate times which later turns into a diffusive behavior \cite{jss11}. Also, for pure DSS, the self-diffusion constant decreases as $\lambda$ is increased. A sudden drop in the diffusion constant at a critical $\lambda$ is observed which is due to the chain formation of DSS \cite{si13}. In both of these studies \cite{jss11,si13} on the pure dipolar fluids, the range of $\lambda$ is limited to $1-7.0$ where the longtime behavior is always diffusive. However, for pure dipolar soft core dumbbells \cite{bmh07} a crossover from longtime diffusive to subdiffusive dynamics at very high dipolar couplings due to network formation. In the present case (the LC-DSS mixture), the subdiffusion at large $\lambda$ can also be attributed to the chain forming tendency of DSS. As shown in Fig.~\ref{fig4}(a), at low $\lambda$ (i.e. at high $\rm T^{*}$), the DSS chains are very small. As $\lambda$ is increased (Fig.~\ref{fig4}(b) and (c)) the size of DSS chains grow due to the strong coupling among the DSS. For a long isotropically distributed DSS chain, its collective motion in a dense environment (provided by the LC matrix) is difficult and, consequently, the DSS show an anomalous transport. 

\begin{figure*}
 \includegraphics[scale=0.8]{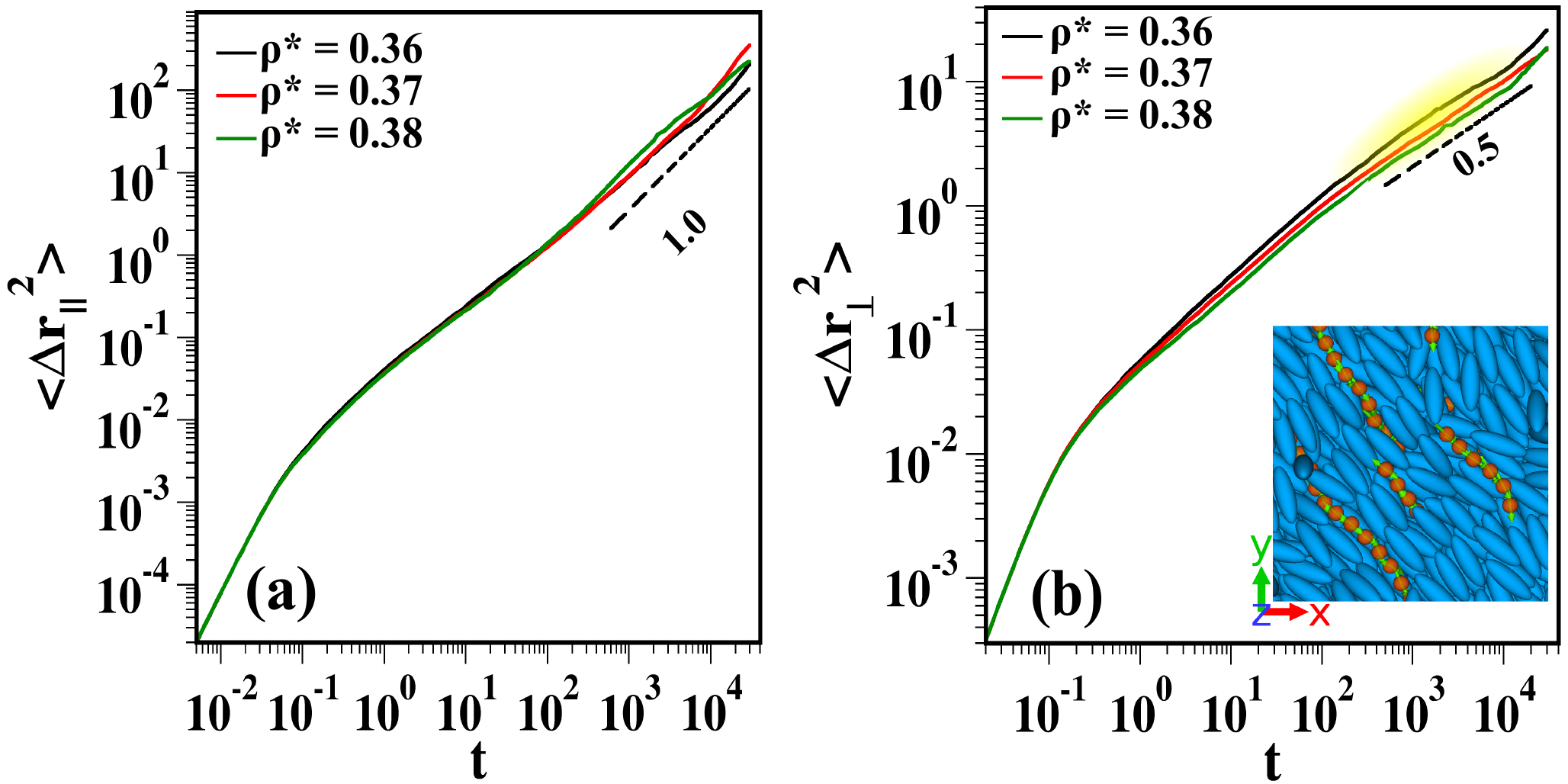}
 \caption{(a). The component of the MSD of the DSS parallel to the $\hat{n}_{s}$ at $\rho^{*} = 0.36, 0.37, 0.38$ and $\lambda = 11.25$. The black dashed lines show a slope 1.0. (b). The component of MSD perpendicular to the $\hat{n}_{s}$ for DSS at the same values of $\rho^{*}$ and $\lambda$ as in (a). The inset shows a top view of the snapshot for $\rho^{*} = 0.37$ and $\lambda = 11.25$. The MSD shows a subdiffusive behavior with exponent 0.5.}
 \label{fig5}
\end{figure*}

In the nematic phase, we fix $\lambda = 11.25$ (i.e. ${\rm T}^{*} = 0.8$) and investigate the MSD for different $\rho^{*}$ in the direction parallel and perpendicular to the $\hat{n}_{s}$. The reason behind such a choice of $\lambda$ is that at this value, the DSS remain subdiffusive during the entire simulation time window and we expect the nematic ordering in the LC matrix to alter the DSS dynamics parallel to the $\hat{n}_{s}$. Fig.~\ref{fig5}(a) shows $\langle \Delta r_{||}^{2}\rangle$ of DSS, which has initial subdiffusive increase (reminiscent of the isotropic phase) which crosses over to a normal diffusive behavior at long times for all considered densities. The $\langle \Delta r_{\perp}^{2}\rangle$ for DSS, plotted in Fig.~\ref{fig5}(b), remains subdiffusive even at long times. The exponent of subdiffusion is $\sim 0.5$ which is slower than the observed exponent in the isotropic regime for the total MSD at the same $\rho^{*}$ and $\lambda$. The inset in Fig.~\ref{fig5}(b) shows the trapping of DSS chains in the cylindrical cavities formed by the LC matrix in the nematic phase.  Due to this confinement their movement perpendicular to $\hat{n}_{s}$ is restricted and the exponent of the subdiffusion decreases with time for all densities. 

Indeed, we would expect a plateau like behavior in the MSD, as the movement of particles is limited by the width of the confining channel. However, despite the strong confinement provided by the LC matrix, the MSD does not saturate to a plateau in the entire simulation time window. Such a behavior arises as the width of the LC channel changes with time due to the diffusion of the LC particles. This leads to very large saturation time scales of the MSD of the DSS. 

\subsubsection{Diffusional behavior of the LC matrix}
\begin{figure*}
 \includegraphics[scale=0.8]{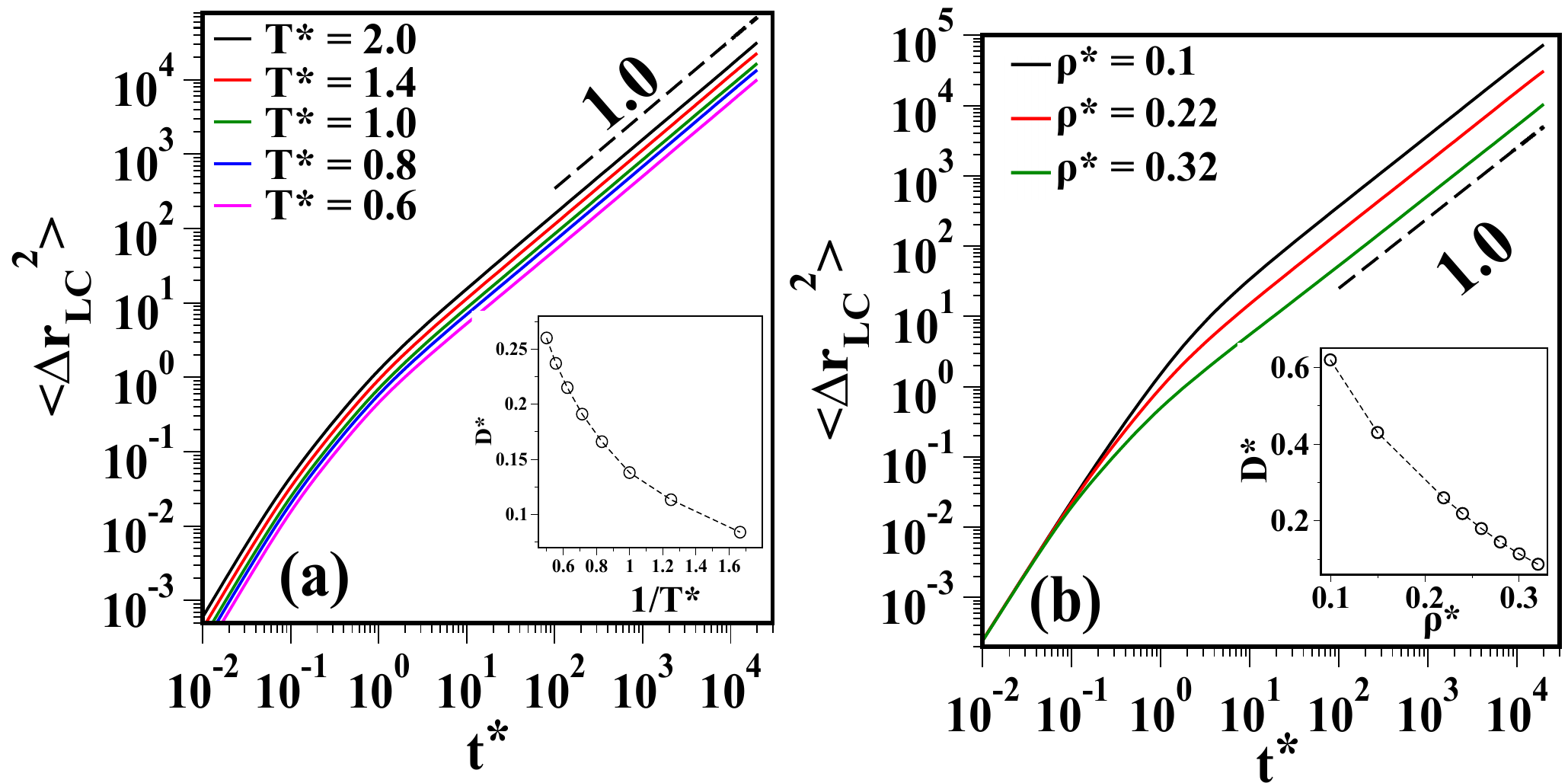}
 \caption{(a). The MSD of the LC matrix in the isotropic phase. (a). The MSD of the LC at fixed $\rho = 0.3$ (along the vertical black dashed line in Fig.~\ref{fig2}) for ${\rm T}^{*} = 2.0, 1.4, 1.0, 0.8, 0.6$. The black dashed line shows the slope 1.0. (b). The MSD of LC matrix at a fixed ${\rm T}^{*} = 0.8$ (along the horizontal black dashed line in Fig.~\ref{fig2}) and $\rho^{*} = 0.1, 0.22, 0.32$. Inset in both the figures shows the variation of self diffusion constant as a function of $1/{\rm T}^{*}$ and $\rho^{*}$ respectively.}
 \label{fig6}
\end{figure*}

In contrast to the DSS, the LC matrix shows normal diffusive dynamics at all densities in the isotropic phase. In Fig.~\ref{fig6}(a), we plot the MSDs of the LC matrix at fixed $\rho^{*}$ and for various $\rm T^{*}$ (along the vertical dashed line in Fig.~\ref{fig2}). Further, Fig.~\ref{fig6}(b) shows the MSDs of LC matrix at  fixed ${\rm T}^{*}$(=0.8) and for different $\rho^{*}$ (along the horizontal dashed line in Fig.~\ref{fig2}). In both of these cases, the longtime behavior of MSDs is normal diffusive for all $\rho^{*}$ and $\rm T^{*}$. Also, the diffusion constant decreases monotonically in both the cases, as can be seen in the insets of the Fig.~\ref{fig6}(a) and (b).

\begin{figure*}
 \centering
 \includegraphics[scale=0.8]{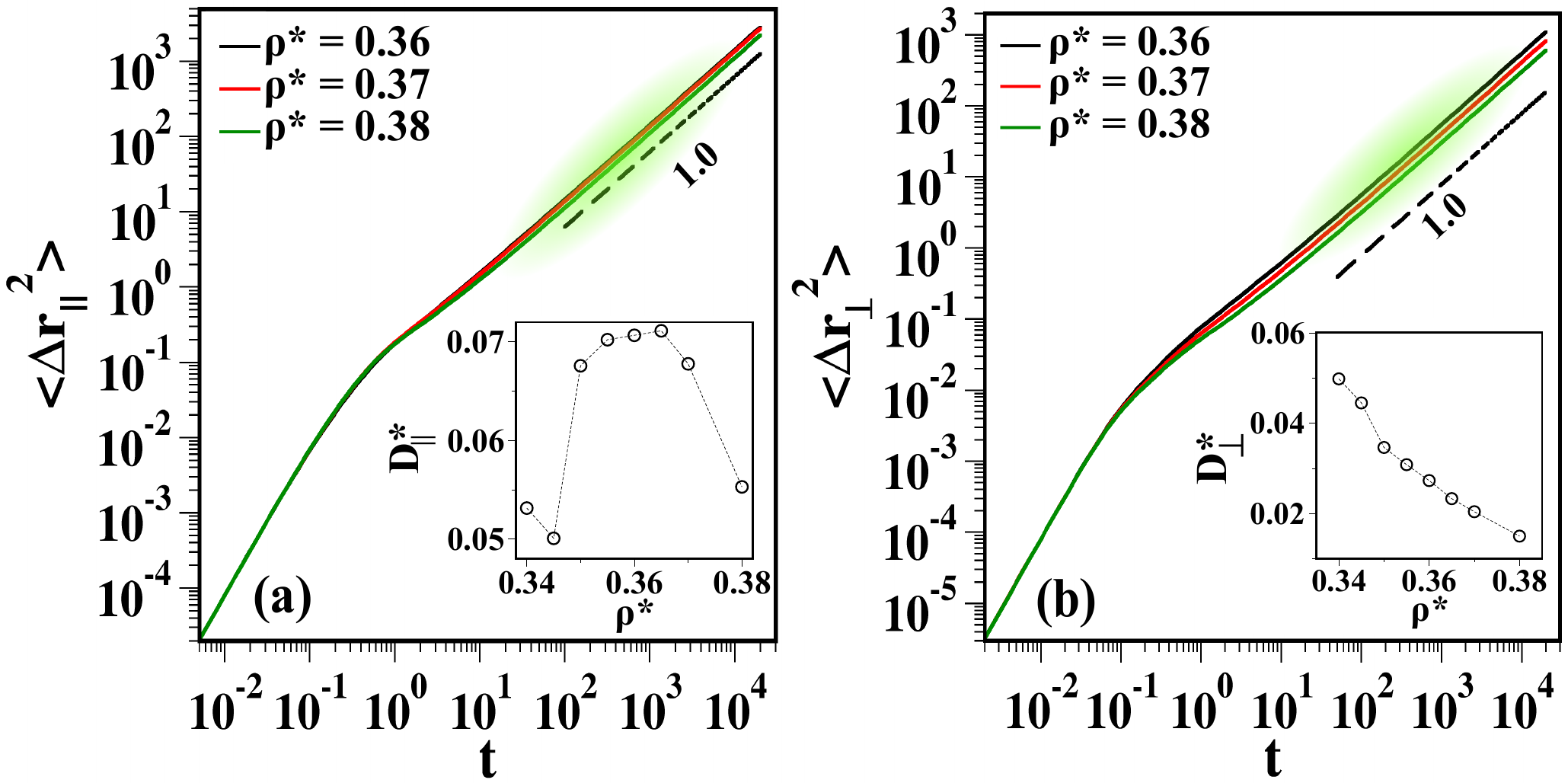}
 \caption{The MSD of the LC matrix in the nematic phase. (a). The component of MSD parallel to $\hat{n}_{e}$ for $\rho^{*} = 0.36, 0.37$ and $0.38$ at ${\rm T}^{*} = 0.8$. (b). The component of MSD perpendicular to $\hat{n}_{e}$ for the same $\rho^{*}$ and ${\rm T}^{*}$ as in (a). The longtime behavior of MSDs in all the three cases is diffusive as shown by black dashed line with slope 1.0. Insets in (a) and (b) show the variation of $D^{*}_{||}$ and $D^{*}_{\perp}$ with $\rho^{*}$.}
 \label{fig7}
\end{figure*}

For the nematic phase, the $\langle \Delta r_{||}^{2}\rangle$ of the LC matrix is plotted in Fig.~\ref{fig7}(a). One observes a normal diffusion at long times for all $\rho^{*}$ in the nematic regime. The inset in Fig.~\ref{fig7}(a) shows the self-diffusion constants $D_{||}$ of the LC matrix parallel to the $\hat{n}_{e}$, which are extracted from the long time behavior of the $\langle \Delta r_{||}^{2}\rangle$. We find that $D_{||}$ shows a non-monotic behavior with increasing $\rho^{*}$, consistent with earlier studies of pure LCs \cite{hl99,mrg92,al90}. The initial fast diffusion of the LC matrix in the nematic regime is attributed to the faster movement of LC particles in the direction parallel to the $\hat{n}_{e}$ \cite{mrg92,al90}. 

Figure~\ref{fig7}(b) shows the $\langle \Delta r_{\perp}^{2}\rangle$ for the LC matrix. The long time behavior of $\langle \Delta r_{\perp}^{2}\rangle$ is also normal diffusive. The $D^{*}_{\perp}$, extracted from $\langle \Delta r_{\perp}^{2}\rangle$, plotted in the inset of Fig.~\ref{fig7}(b), decreases monotonically with increasing $\rho^{*}$ which is consistent with the earlier observations \cite{mrg92,al90}. 

\subsection{Velocity autocorrelation functions}
\label{vacf}
To explore the local environment of particles, we study the normalized VACF which is defined as $C_{v}(t) = \langle \bm{v}(t_{0})\cdot\bm{v}(t+t_{0})\rangle/\langle \bm{v}(0)\cdot\bm{v}(0)\rangle$ \cite{at06}. The VACF resolved in the direction parallel and perpendicular to the $\hat{n}_{s}$, $C_{v_{||, \perp}}(t)$, is defined similar to the $C_{v}(t)$, where only the respective components of the velocity are considered.

\begin{figure*}
 \centering
 \includegraphics[scale=0.8]{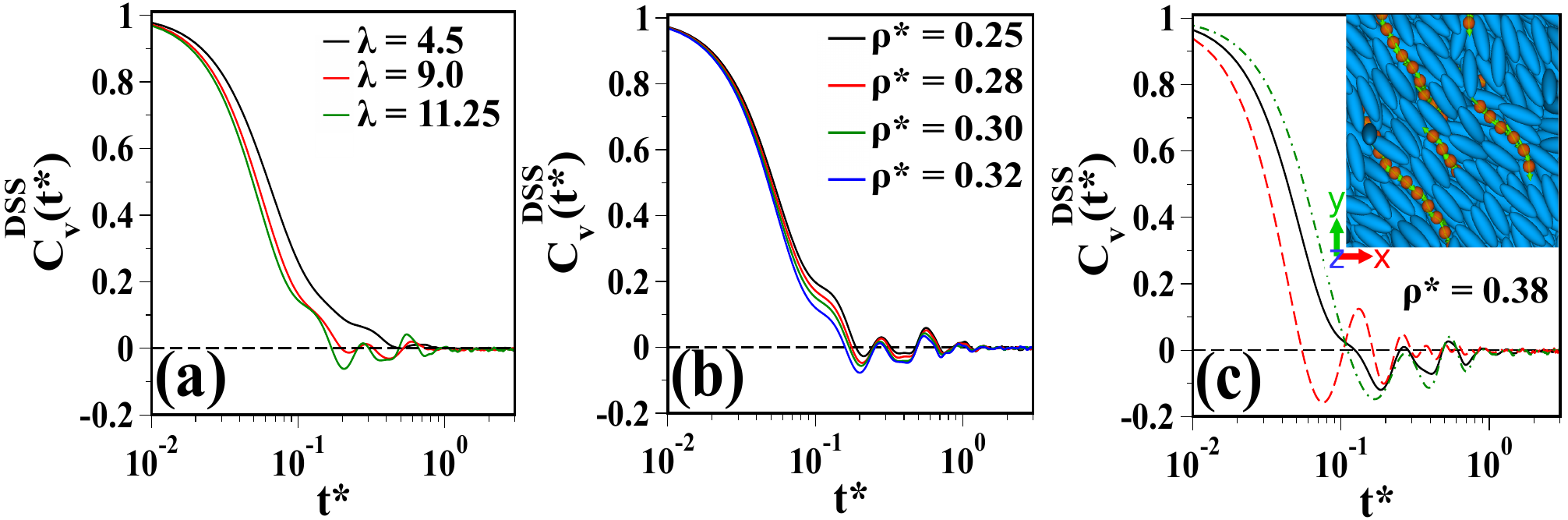}
 \caption{The VACF of the DSS in the isotropic phase (a) for $\lambda = 4.5, 6.43, 11.25$ at a fixed $\rho^{*} = 0.3$ (along the vertical black dashed line in Fig.~\ref{fig2}) and (b) for $\rho^{*} = 0.25, 0.28, 0.3, 0.32$ at $\lambda = 11.25$ (along the horizontal black dashed line in Fig.~\ref{fig2}). (c) The VACF for DSS in the nematic regime at $\rho^{*} = 0.38$ and $\lambda = 11.25$. The red dashed and green dot-dashed lines represent the function $C_{v_{||}}(t)$ and $C_{v_{\perp}}(t)$. The inset shows a top view of the snapshot for $\rho^{*} = 0.37$ and $\lambda = 11.25$.}
 \label{fig8}
\end{figure*}

Results for the VACFs for the DSS in the isotropic regime are plotted in the Fig.~\ref{fig8}(a) for $\lambda = 4.5, 9.0$ and $11.25$ at $\rho^{*} = 0.3$ (along the vertical dashed line in Fig.~\ref{fig2}). At low $\lambda$, the VACF decay smoothly without any oscillations. This indicates that the particles move essentially as unbounded objects, that is, chains are not yet formed. As $\lambda$ is increased oscillations in the VACF are observed at short time scales which we attribute to chain formation. The oscillatory behavior at short times appears due to the rattling of DSS particles in the chain and indicates its ``caging" due to strong dipolar coupling. 

In the case of pure dipolar fluids in the isotropic phase at comparable $\lambda$ \cite{wp93}, such an oscillatory behavior is not observed (cf. red curve in Fig.~\ref{fig8}(a)). Also, for pure dipolar dumbbells, the oscillatory behavior in VACF is observed only at a very large $\lambda$ where dipolar particles form a percolating network \cite{bmh07}. In our case, we observe the oscillatory behavior at $\lambda \sim 6.43$ where a percolating network is not yet expected (see, e.g., Fig~\ref{fig4}(b), where long but not system spanning chains are visible at $\lambda = 6.43$). We understand the difference between the present mixed system and the pure system as follows: for pure dipolar fluids with $\lambda = 6.66$ in the isotropic phase \cite{wp93}, the dipolar chains are randomly distributed and their collective motion is not hindered. Therefore, an oscillatory behavior does not occur in the VACFs of the DSS. In the present system, the DSS chains experience a dense environment provided by the LC matrix, which is reflected in the form of oscillations in the VACFs of the DSS.

It should also be noted that in Fig.~\ref{fig8}(a) at small $\lambda$, where DSS chains are much shorter (e.g., Fig.~{fig4}(a)), there is no negative lobe present in the VACF, which suggests that the short DSS chains or single DSS particles do not feel the dense environment. As the size of the chain grows upon increase of $\lambda$ (for a fixed $\rho^{*}$), the role the LC matrix becomes more evident in modifying the dynamics. Fig.~\ref{fig8}(b) shows the VACF of DSS at a fixed $\rm T^{*} = 0.8$ and various $\rho^{*}$ (along the horizontal black dashed line in Fig.~\ref{fig2}). Here, the oscillatory behavior can be seen at all considered $\rho^{*}$.

In the nematic phase, the $C_{v_{||}}(t)$ for the DSS chains, plotted in Fig.~\ref{fig8}(c) (shown by red dashed line) decays faster than $C_{v_{\perp}}(t)$ (represented by green dot-dashed line). The oscillations after the smooth initial decay are present in both the components. This is contrary to earlier observations for pure DSS where oscillations in the $C_{v_{\perp}}$ are not found \cite{wp93}. The oscillations in the parallel component arise due to the strong dipolar interactions which force the particles to remain in the chain. In the present case, the oscillations in the perpendicular component are the result of  the confinement of the DSS chains in a narrow cylindrical channel formed by the LC matrix as shown in the inset of Fig.~\ref{fig8}(b). Such an oscillatory behavior is peculiar to the dipolar fluids confined in a narrow spaces \cite{fk06}.

\begin{figure*}
 \centering
 \includegraphics[scale=0.8]{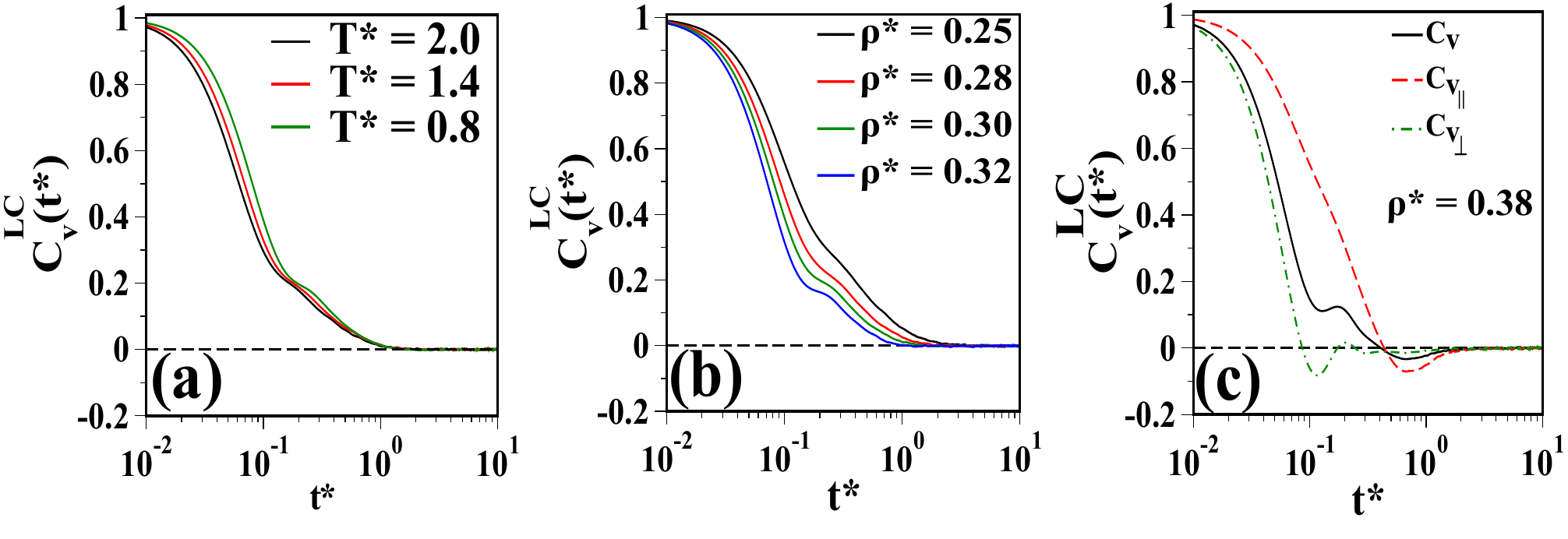}
 \caption{The VACF of the LC matrix in the isotropic phase (a) for ${\rm T}^{*} = 2.0, 1.4, 0.8$ at fixed $\rho^{*} = 0.3$ (along the vertical black dashed line in Fig.~\ref{fig2}) and (b) for $\rho^{*} = 0.25, 0.28, 0.3, 0.32$ at ${\rm T}^{*} = 0.8$ (along the horizontal black dashed line in Fig.~\ref{fig2}). (c). The VACF for LC in the nematic regime at the $\rho^{*} = 0.38$ and ${\rm T}^{*} = 0.8$. The red dashed and green dot-dashed lines represent the $C_{v_{||}}(t)$ and $C_{v_{\perp}}(t)$. Inset shows the $C_{v_{||}}(t)$ for $\rho^{*} = 0.34, 0.35, 0.36, 0.38$.}
 \label{fig9}
\end{figure*}

We now turn to the VACF for the LC matrix. Results for the isotropic phase along the two dashed lines shown in Fig.~\ref{fig2} are plotted in  Fig.~\ref{fig9}. Similar to the earlier observations, Fig.~\ref{fig9}(a) shows that the VACFs do not show any negative lobe for the whole range of $\rm T^{*}$ considered in the isotropic phase at $\rho^{*} = 0.3$ (along the vertical black dashed line in Fig.~\ref{fig2}). At lower $\rm T^{*}$, a plateau appears which later converts into a minimum and a maximum \cite{mrg92}. Similar behavior is observed (Fig.~\ref{fig9}(b)) at fixed $\rm T^{*} = 0.8$ and various $\rho^{*}$ (along the horizontal black dashed line in Fig.~\ref{fig2}).

Finally, the VACF of the LCs in the nematic regime for $\rho^{*} = 0.38$ and $\rm T^{*} = 0.8$ is plotted in the Fig.~\ref{fig9}(c). The ``shoulder'' (the minimum  and the subsequent maximum), which was observed at higher isotropic densities (cf. Fig.~\ref{fig9}(b)), becomes more pronounced in the nematic regime. Further, the decay of $C_{v_{||}}(t)$ is slower than that of $C_{v_{\perp}}(t)$ consistent with the larger value of $D^{*}_{||}$. We observe a small negative region in $C_{v_{||}}(t)$ which is an indicator of the high density of the mixture. As discussed in earlier studies \cite{mrg92}, the oscillatory behavior in $C_{v_{\perp}}(t)$ (shown by green dot-dashed line in Fig.~\ref{fig9}(c)) is due to the periodic rebounds of the LC particles in the direction perpendicular to the $\hat{n}_{e}$.

\subsubsection{Signature of the subdiffusion in the VACF}
So far we have analyzed the short-time behavior of the VACF of the DSS in order to investigate the local environment around the DSS. We, now, show that the subdiffusive behavior, observed in the MSD of the DSS, is also reflected in the VACF. To this end, we start with the Green-Kubo relation in three dimensions \cite{fs01},
\begin{eqnarray}
\label{tdd}
D = \frac{1}{3}\int_{0}^{\infty}\langle \bm v(t^{\prime})\cdot \bm v(0)\rangle dt^{\prime}.
\end{eqnarray}
Here, $t^{\prime} = t - t_{0}$ is the time elapsed from the time origin taken at $t_{0}$. The VACF is related to the MSD as \cite{hf13,fs01},
\begin{eqnarray}
\label{vcfmsd}
\frac{1}{3}\langle \bm v(t)\cdot \bm v(0)\rangle
	   = \frac{1}{6}\frac{d^{2}}{dt^{2}} \langle {\Delta r}^{2}(t)\rangle	     
\end{eqnarray}
Inserting Eq.~(\ref{vcfmsd}) into Eq.~(\ref{tdd}) we obtain,
\begin{eqnarray}
\label{tdmsd}
D = \frac{1}{3}\int_{0}^{\infty}\langle \bm v(t^{\prime})\cdot \bm v(0)\rangle dt^{\prime} = \frac{1}{6}\frac{d}{dt^{\prime}} \langle {\Delta r}^{2}(t^{\prime})\rangle\mid_{t^{\prime} \to \infty}.
\end{eqnarray} 
In simulations, we need to consider a ``sufficiently large" time interval $\tau$ up to which numerical integration is performed in order to calculate the diffusion coefficient. Starting from this, we define a time (interval) dependent diffusion coefficient $D(\tau)$ by rewriting Eq.~(\ref{tdmsd}) as,
\begin{eqnarray}
\label{tddss}
D(\tau) = \frac{1}{3}\int_{0}^{\tau}\langle \bm v(t^{\prime})\cdot \bm v(0)\rangle dt^{\prime} = \frac{1}{6}\frac{d}{dt^{\prime}} \langle {\Delta r}^{2}(t^{\prime})\rangle\mid_{t^{\prime} \to \tau}.
\end{eqnarray}
Therefore, for a diffusive process, in the limit of large $\tau$, $D(\tau)$ saturates to the diffusion coefficient obtained via the Stokes-Einstein relation $\langle \Delta r^{2}(t^{\prime}) = 6Dt^{\prime} \rangle $. Whereas, in the cases where the longtime behavior of the MSD is subdiffusive with an exponent $\alpha$, $D(\tau)$ should show a  power-law decay as a function of $\tau$ with an exponent ($\alpha -1$).  

\begin{figure*}
 \centering
 \includegraphics[scale=0.6]{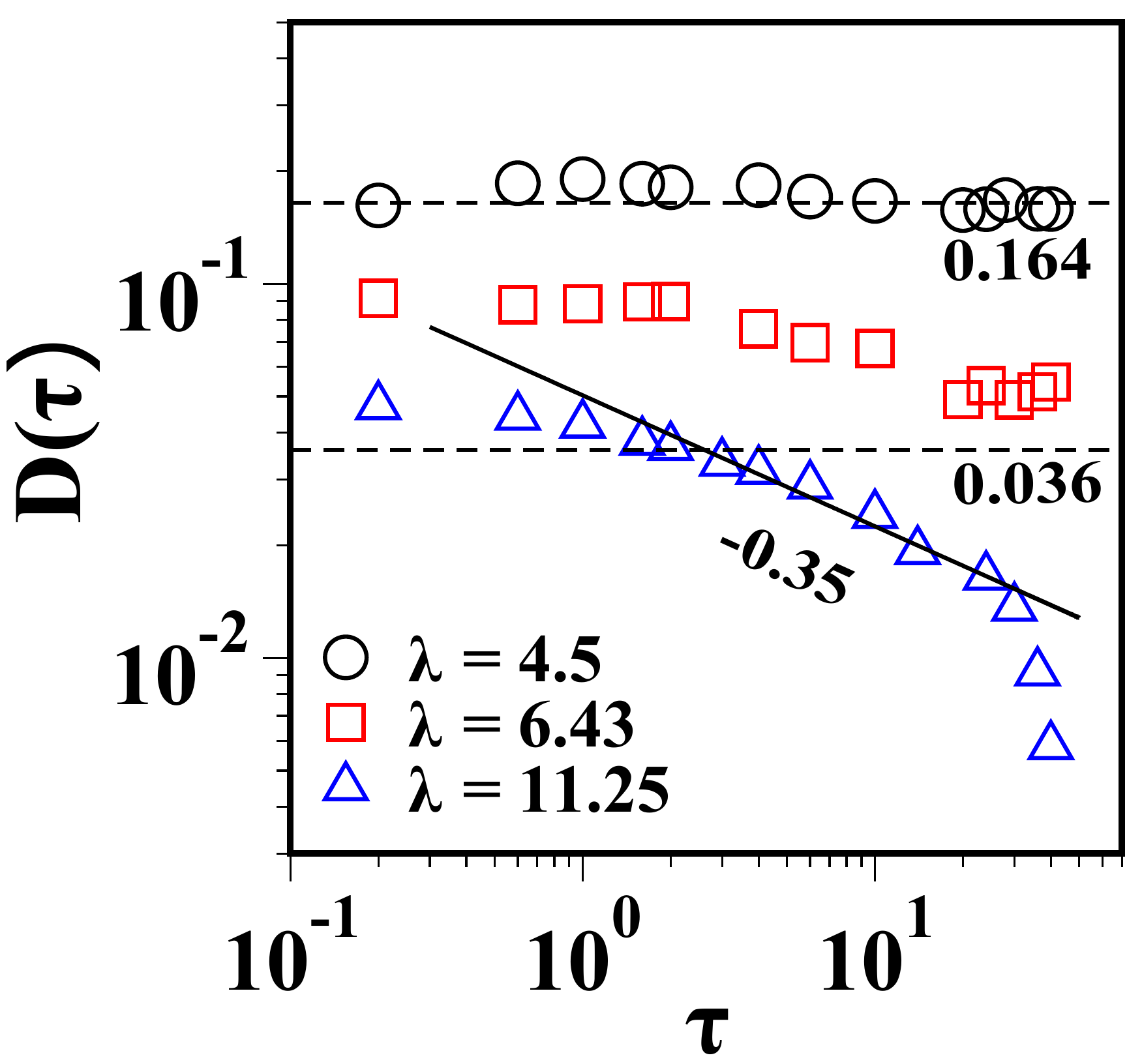}
 \caption{Variation of the time dependent diffusion coefficient $D(\tau)$ calculated from the VACF of the DSS as a function of time interval $\tau$ for $\lambda = 4.5, 6.43$ and $11.25$ at $\rho^{*} = 0.3$. The black dashed lines show the value of $D$ obtained from the MSD of the DSS for $\lambda = 4.5$ and $6.43$. The solid black line represents the slope -0.35.}
 \label{fig12}
\end{figure*}

In Fig.~\ref{fig12}, we plot $D(\tau)$ as function of $\tau$ calculated from the VACF of the DSS for $\lambda = 4.5, 6.43$ and $11.25$ at $\rho^{*} = 0.3$ (the green filled squares marked on the vertical dashed line in Fig.~\ref{fig2}). Clearly, for $\lambda = 4.5, 6.43$, $D(\tau)$ saturates to $D$ which is calculated from the longtime behavior of the corresponding MSD of the DSS (values of $D$ are shown by dashed black lines in Fig.~\ref{fig12}). Further, for $\lambda = 11.25$, $D(\tau)$ shows a power-law decay with an exponent -0.35, which is consistent with the exponent of subdiffusion obtained from the corresponding longtime behavior of the MSD of the DSS. Thus, our numerical results confirm the relation between the VACF and the MSD in the longtime limit. The slight discrepancy in the saturated value of the $D(\tau)$ for $\lambda = 6.43$ and the sudden drop in the $D(\tau)$ for $\lambda = 11.25$ at long times may be due to the lack of statistical averaging. 
\section{Summary and outlook}
\label{con}
To summarize, we have presented a detailed MD simulation study on the translational dynamics in a LC-DSS mixture, in which both the species have comparable sizes. Our main finding is that in such a mixture, the DSS show a crossover from normal to anomalous translational dynamics in the isotropoic regime as $\lambda$ is increased. At small $\lambda$ the DSS chain lengths are shorter and, therefore, they can easily diffuse inside the LC matrix. However, as $\lambda$ is increased, the lengths of the DSS chains  grow and eventually their translational dynamics become subdiffusive. In this sense, the orientationally disordered LC matrix provides a complex environment which strongly influences the diffusion of host particles.

In the nematic regime, the LC particles, due to their cooperative  movement and reorientation in the direction parallel to the $\hat{n}_{e}$, form a cylindrical channel and force the DSS chains to align along the channel. As a consequence, the DSS diffuse normally parallel to the $\hat{n}_{e}$ while remain subdiffusive in the perpendicular direction. The anisotropic translational dynamics of DSS in narrow slit pores is well studied, and enhanced diffusion of DSS parallel to the channel has been observed \cite{fk06}. The DSS chains in the present case show similar behavior. 

The strong confinement imposed by the LC matrix is also reflected in the VACFs of the DSS, which show an oscillatory behavior not only in the isotropic regime but also in the nematic regime. The VACFs of the LC matrix show a minimum and maximum at short times in the isotropic regime. In the nematic regime, only the perpendicular component of the VACFs of the LC matrix shows an oscillatory behavior while the parallel component decays smoothly at high nemtic densities.

In this work, we have demonstrated that the LC matrix strongly modifies the translational dynamics of the DSS whereas the LC dynamics remains essentially unaffected. A further study of the rotational dynamics of such mixtures is necessary in order to develop a better understanding of the intrinsic timescles. Also, it will be interesting to explore the effect of external magnetic field on the dynamics of different components of the mixture, as it has been experimentally shown that a weak magnetic field affects the ordering in these systems \cite{mespkk16}.

We believe that the present study will be very helpful in understanding the rheology of ferronematics as the interplay of intrinsic and shear induced timescales give rise to intriguing behavior in these complex fluids. Walk in these directions is under way.

\section{Acknowledgements}
We gratefully acknowledge funding support from the Deutsche Forschungsgemainschaft (DFG) via the priority program SPP 1681. GPS would like to thank J\"{u}rgen Horbach, Nima H. Siboni and Sascha Gerloff for fruitful discussions.














\end{document}